\input harvmac
\input epsf
\input amssym
%
%
\noblackbox
\newcount\figno
\figno=0
\def\fig#1#2#3{
\par\begingroup\parindent=0pt\leftskip=1cm\rightskip=1cm\parindent=0pt
\baselineskip=11pt
\global\advance\figno by 1
\midinsert
\epsfxsize=#3
\centerline{\epsfbox{#2}}
\vskip -21pt
{\bf Fig.\ \the\figno: } #1\par
\endinsert\endgroup\par
}
\def\figlabel#1{\xdef#1{\the\figno}}
\def\encadremath#1{\vbox{\hrule\hbox{\vrule\kern8pt\vbox{\kern8pt
\hbox{$\displaystyle #1$}\kern8pt}
\kern8pt\vrule}\hrule}}

\def\frac#1#2{{#1 \over #2}}

\def\semi{\subset\kern-1em\times\;}
\def\bar#1{\overline{#1}}
\def\sqr#1#2{{\vcenter{\vbox{\hrule height.#2pt
\hbox{\vrule width.#2pt height#1pt \kern#1pt \vrule width.#2pt}
\hrule height.#2pt}}}}

\def\ad{\bar a}

\def\go{g^{(0)} }

%

%


\lref\masakiref{
  N.~Iizuka and M.~Shigemori,
  ``A Note on D1-D5-J System and 5D Small Black Ring,''
  arXiv:hep-th/0506215.
}

\lref\KLatt{
  P.~Kraus and F.~Larsen,
  ``Attractors and black rings,''
  Phys.\ Rev.\ D {\bf 72}, 024010 (2005)
  [arXiv:hep-th/0503219].
}

\lref\ChamseddinePI{
  A.~H.~Chamseddine, S.~Ferrara, G.~W.~Gibbons and R.~Kallosh,
  ``Enhancement of supersymmetry near 5d black hole horizon,''
  Phys.\ Rev.\ D {\bf 55}, 3647 (1997)
  [arXiv:hep-th/9610155].
}

\lref\StromBTZ{ A.~Strominger,
 ``Black hole entropy from near-horizon microstates'',
JHEP {\bf 9802}, 009 (1998); [arXiv:hep-th/9712251];
V.~Balasubramanian and F.~Larsen, ``Near horizon geometry and black
holes in four dimensions'', Nucl.\ Phys.\ B {\bf 528}, 229 (1998);
[arXiv:hep-th/9802198].
}

\lref\BalasubramanianEE{
  V.~Balasubramanian and F.~Larsen,
  Nucl.\ Phys.\  B {\bf 528}, 229 (1998)
  [arXiv:hep-th/9802198].
}

\lref\MSW{ J.~M.~Maldacena, A.~Strominger and E.~Witten, ``Black
hole entropy in M-theory'', JHEP {\bf 9712}, 002 (1997);
[arXiv:hep-th/9711053]. }

\lref\HMM{ J.~A.~Harvey, R.~Minasian and G.~W.~Moore, ``Non-abelian
tensor-multiplet anomalies,''
 JHEP {\bf 9809}, 004 (1998)
  [arXiv:hep-th/9808060].
}

\lref\antRRRR{  I.~Antoniadis, S.~Ferrara, R.~Minasian and
K.~S.~Narain,
  ``R**4 couplings in M- and type II theories on Calabi-Yau spaces,''
  Nucl.\ Phys.\ B {\bf 507}, 571 (1997)
  [arXiv:hep-th/9707013].
 }

\lref\WittenMfive{ E.~Witten,
  ``Five-brane effective action in M-theory,''
  J.\ Geom.\ Phys.\  {\bf 22}, 103 (1997)
  [arXiv:hep-th/9610234].
}

\lref\wittenAdS{ E.~Witten,
  ``Anti-de Sitter space and holography,''
  Adv.\ Theor.\ Math.\ Phys.\  {\bf 2}, 253 (1998)
  [arXiv:hep-th/9802150].
  }

\lref\brownhen{  J.~D.~Brown and M.~Henneaux,
 ``Central Charges In The Canonical Realization Of Asymptotic Symmetries: An
  Example From Three-Dimensional Gravity,''
  Commun.\ Math.\ Phys.\  {\bf 104}, 207 (1986).
  }

\lref\wald{
  R.~M.~Wald,
  ``Black hole entropy is the Noether charge,''
  Phys.\ Rev.\ D {\bf 48}, 3427 (1993)
  [arXiv:gr-qc/9307038].
R.~Wald, Phys.\ Rev.\ D {\bf 48} R3427 (1993);
   V.~Iyer and R.~M.~Wald,
  ``Some properties of Noether charge and a proposal for dynamical black hole
  entropy,''
  Phys.\ Rev.\ D {\bf 50}, 846 (1994)
  [arXiv:gr-qc/9403028].
 ``A Comparison of Noether charge and Euclidean methods for computing the
  entropy of stationary black holes,''
  Phys.\ Rev.\ D {\bf 52}, 4430 (1995)
  [arXiv:gr-qc/9503052].
}

\lref\senrescaled{A.~Sen,
  ``How does a fundamental string stretch its horizon?,''
  JHEP {\bf 0505}, 059 (2005)
  [arXiv:hep-th/0411255];
   ``Black holes, elementary strings and holomorphic anomaly,''
     JHEP {\bf 0507}, 063 (2005)
  [arXiv:hep-th/0502126];
   ``Stretching the horizon of a higher dimensional small black hole,''
    JHEP {\bf 0507}, 073 (2005)
  [arXiv:hep-th/0505122];
   ``Entropy function for heterotic black holes,''
      JHEP {\bf 0603}, 008 (2006)
  [arXiv:hep-th/0508042]
;
  B.~Sahoo and A.~Sen,
  ``alpha' corrections to extremal dyonic black holes in heterotic string
  theory,''
  JHEP {\bf 0701}, 010 (2007)
  [arXiv:hep-th/0608182].
}

\lref\SenWA{
  A.~Sen,
  ``Black hole entropy function and the attractor mechanism in higher
  derivative gravity,''
  JHEP {\bf 0509}, 038 (2005)
  [arXiv:hep-th/0506177].
}

\lref\saidasoda{
  H.~Saida and J.~Soda,
  ``Statistical entropy of BTZ black hole in higher curvature gravity,''
  Phys.\ Lett.\ B {\bf 471}, 358 (2000)
  [arXiv:gr-qc/9909061].
}

\lref\attract{ S.~Ferrara, R.~Kallosh and A.~Strominger, ``N=2
extremal black holes'', Phys.\ Rev.\ D {\bf 52}, 5412 (1995),
[arXiv:hep-th/9508072];
 A.~Strominger,
 ``Macroscopic Entropy of $N=2$ Extremal Black Holes'',
 Phys.\ Lett.\ B {\bf 383}, 39 (1996),
[arXiv:hep-th/9602111];
S.~Ferrara and R.~Kallosh, ``Supersymmetry and Attractors'', Phys.\
Rev.\ D {\bf 54}, 1514 (1996), [arXiv:hep-th/9602136];
``Universality of Supersymmetric Attractors'', Phys.\ Rev.\ D {\bf
54}, 1525 (1996), [arXiv:hep-th/9603090];
R.~Kallosh, A.~Rajaraman and W.~K.~Wong, ``Supersymmetric rotating
black holes and attractors'', Phys.\ Rev.\ D {\bf 55}, 3246 (1997),
[arXiv:hep-th/9611094];
A~Chou, R.~Kallosh, J.~Rahmfeld, S.~J.~Rey, M.~Shmakova and
W.~K.~Wong, ``Critical points and phase transitions in 5d
compactifications of M-theory''. Nucl.\ Phys.\ B {\bf 508}, 147
(1997); [arXiv:hep-th/9704142].
}

\lref\moore{G.~W.~Moore,``Attractors and arithmetic'',
[arXiv:hep-th/9807056];
``Arithmetic and attractors'', [arXiv:hep-th/9807087];
``Les Houches lectures on strings and arithmetic'',
[arXiv:hep-th/0401049];
B.~R.~Greene and C.~I.~Lazaroiu, ``Collapsing D-branes in Calabi-Yau
moduli space. I'', Nucl.\ Phys.\ B {\bf 604}, 181 (2001),
[arXiv:hep-th/0001025]. }

\lref\ChamseddinePI{
  A.~H.~Chamseddine, S.~Ferrara, G.~W.~Gibbons and R.~Kallosh,
  ``Enhancement of supersymmetry near 5d black hole horizon,''
  Phys.\ Rev.\ D {\bf 55}, 3647 (1997)
  [arXiv:hep-th/9610155].
}

\lref\denef{  
F.~Denef,``Supergravity flows and D-brane stability'', JHEP {\bf
0008}, 050 (2000), [arXiv:hep-th/0005049];
``On the correspondence between D-branes and stationary supergravity
 solutions of type II Calabi-Yau compactifications'',
[arXiv:hep-th/0010222];
``(Dis)assembling special Lagrangians'', [arXiv:hep-th/0107152].
  B.~Bates and F.~Denef,
   ``Exact solutions for supersymmetric stationary black hole composites,''
  arXiv:hep-th/0304094.
}

\lref\OSV{H.~Ooguri, A.~Strominger and C.~Vafa, ``Black hole
attractors and the topological string'', Phys.\ Rev.\ D {\bf 70},
106007 (2004), [arXiv:hep-th/0405146];
}

\lref\moreOSV{ J.~de Boer, M.~C.~N.~Cheng, R.~Dijkgraaf, J.~Manschot
and E.~Verlinde, ``A farey tail for attractor black holes,'' JHEP
{\bf 0611}, 024 (2006) [arXiv:hep-th/0608059].
}

\lref\DabholkarYR{
  A.~Dabholkar,
  ``Exact counting of black hole microstates,''
  Phys.\ Rev.\ Lett.\  {\bf 94}, 241301 (2005)
  [arXiv:hep-th/0409148].
}

 \lref\DDMP{
A.~Dabholkar, F.~Denef, G.~W.~Moore and B.~Pioline, ``Exact and
asymptotic degeneracies of small black holes'',
[arXiv:hep-th/0502157];  ``Precision counting of small black
holes,''
  JHEP {\bf 0510}, 096 (2005)
  [arXiv:hep-th/0507014].
}

\lref\curvcorr{A.~Dabholkar, ``Exact counting of black hole
microstates", [arXiv:hep-th/0409148],
A.~Dabholkar, R.~Kallosh and A.~Maloney, ``A stringy cloak for a
classical singularity'', JHEP {\bf 0412}, 059 (2004),
[arXiv:hep-th/0410076].
} \lref\bkmicro{
 I.~Bena and P.~Kraus,
 ``Microscopic description of black rings in AdS/CFT'',
JHEP {\bf 0412}, 070 (2004)
  [arXiv:hep-th/0408186].
} \lref\cgms{ M.~Cyrier, M.~Guica, D.~Mateos and A.~Strominger,
``Microscopic entropy of the black ring'', [arXiv:hep-th/0411187].
}

\lref\CardosoFP{
  K.~Behrndt, G.~Lopes Cardoso, B.~de Wit, D.~Lust, T.~Mohaupt and
  W.~A.~Sabra,
  ``Higher-order black-hole solutions in N = 2 supergravity and
  Calabi-Yau
  string backgrounds,''
  Phys.\ Lett.\ B {\bf 429}, 289 (1998)
  [arXiv:hep-th/9801081];
[arXiv:hep-th/0012232]; ``Stationary BPS solutions in N = 2
supergravity with $R^2 $ interactions'', JHEP {\bf 0012}, 019 (2000)
[arXiv:hep-th/0009234];
  ``Macroscopic entropy formulae and non-holomorphic corrections for
  supersymmetric black holes'',
  Nucl.\ Phys.\ B {\bf 567}, 87 (2000)
  [arXiv:hep-th/9906094];
  ``Corrections to macroscopic supersymmetric black-hole entropy'',
  Phys.\ Lett.\ B {\bf 451}, 309 (1999)
  [arXiv:hep-th/9812082].
}

\lref\hensken{  M.~Henningson and K.~Skenderis,
  ``The holographic Weyl anomaly,''
  JHEP {\bf 9807}, 023 (1998)
  [arXiv:hep-th/9806087].
  }

\lref\balkraus{  V.~Balasubramanian and P.~Kraus,
  ``A stress tensor for anti-de Sitter gravity,''
  Commun.\ Math.\ Phys.\  {\bf 208}, 413 (1999)
  [arXiv:hep-th/9902121].
  }

\lref\HanakiPJ{
  K.~Hanaki, K.~Ohashi and Y.~Tachikawa,
  ``Supersymmetric completion of an R**2 term in five-dimensional
  supergravity,''
  [arXiv:hep-th/0611329].
}

\lref\us{
  P.~Kraus and F.~Larsen,
  ``Microscopic black hole entropy in theories with higher derivatives,''
  JHEP {\bf 0509}, 034 (2005)
  [arXiv:hep-th/0506176].
}

\lref\SahooVZ{
  B.~Sahoo and A.~Sen,
  ``BTZ black hole with Chern-Simons and higher derivative terms,''
  JHEP {\bf 0607}, 008 (2006)
  [arXiv:hep-th/0601228].
}

\lref\GutowskiYV{
  J.~B.~Gutowski and H.~S.~Reall,
  ``General supersymmetric AdS(5) black holes,''
  JHEP {\bf 0404}, 048 (2004)
  [arXiv:hep-th/0401129].
}

\lref\LarsenXM{
  F.~Larsen,
  ``The attractor mechanism in five dimensions,''
  [arXiv:hep-th/0608191].
}

\lref\MaldacenaBW{
  J.~M.~Maldacena and A.~Strominger,
  ``AdS(3) black holes and a stringy exclusion principle,''
  JHEP {\bf 9812}, 005 (1998)
  [arXiv:hep-th/9804085].
}

\lref\BanadosGQ{
  M.~Banados, M.~Henneaux, C.~Teitelboim and J.~Zanelli,
  ``Geometry of the (2+1) black hole,''
  Phys.Rev.D {\bf 66}, 010001 (2002)
  [arXiv:gr-qc/9302012].
}

\lref\StromingerYG{
  A.~Strominger,
  ``AdS(2) quantum gravity and string theory,''
  JHEP {\bf 9901}, 007 (1999)
  [arXiv:hep-th/9809027].
}

\lref\MohauptMJ{
  T.~Mohaupt,
  ``Black hole entropy, special geometry and strings,''
  Phys.Rev.D {\bf 66}, 010001 (2002)
  arXiv:hep-th/0007195.
}

\lref\BehrndtAC{
  K.~Behrndt, G.~L.~Cardoso and S.~Mahapatra,
  ``On the relation between BPS solutions in 4D and 5D,''
  Int.\ J.\ Mod.\ Phys.\ D {\bf 15}, 1603 (2006).
}

\lref\KrausZM{
  P.~Kraus and F.~Larsen,
  ``Holographic gravitational anomalies,''
  JHEP {\bf 0601}, 022 (2006)
  [arXiv:hep-th/0508218].
}

\lref\GaiottoNS{
  D.~Gaiotto, A.~Strominger and X.~Yin,
  ``From AdS(3)/CFT(2) to black holes / topological strings,''
  [arXiv:hep-th/0602046].
}

\lref\MohauptMJ{
  T.~Mohaupt,
  ``Black hole entropy, special geometry and strings,''
  Fortsch.\ Phys.\  {\bf 49}, 3 (2001)
  [arXiv:hep-th/0007195].
}

\lref\PiolineNI{
  B.~Pioline,
  ``Lectures on on black holes, topological strings and quantum attractors,''
  Class.\ Quant.\ Grav.\  {\bf 23}, S981 (2006)
  [arXiv:hep-th/0607227].
}

\lref\KrausWN{
  P.~Kraus,
  ``Lectures on black holes and the AdS(3)/CFT(2) correspondence,''
  [arXiv:hep-th/0609074].
}

\lref\DabholkarTB{
  A.~Dabholkar, A.~Sen and S.~P.~Trivedi,
  ``Black hole microstates and attractor without supersymmetry,''
  JHEP {\bf 0701}, 096 (2007)
  [arXiv:hep-th/0611143].
}

\lref\rings{
  I.~Bena and P.~Kraus,
  ``R**2 corrections to black ring entropy,''
  arXiv:hep-th/0506015;
  N.~Iizuka and M.~Shigemori,
  ``A note on D1-D5-J system and 5D small black ring,''
  JHEP {\bf 0508}, 100 (2005)
  [arXiv:hep-th/0506215]; A.~Dabholkar, N.~Iizuka, A.~Iqubal and M.~Shigemori,
  ``Precision microstate counting of small black rings,''
  Phys.\ Rev.\ Lett.\  {\bf 96}, 071601 (2006)
  [arXiv:hep-th/0511120]; A.~Dabholkar, N.~Iizuka, A.~Iqubal, A.~Sen and M.~Shigemori,
  ``Spinning strings as small black rings,''
 [arXiv:hep-th/0611166].
}

\lref\TachikawaSZ{
  Y.~Tachikawa,
  ``Black hole entropy in the presence of Chern-Simons terms,''
  Class.\ Quant.\ Grav.\  {\bf 24}, 737 (2007)
  [arXiv:hep-th/0611141].
}

\lref\KugoHN{
  T.~Kugo and K.~Ohashi,
  ``Supergravity tensor calculus in 5D from 6D,''
  Prog.\ Theor.\ Phys.\  {\bf 104}, 835 (2000)
  [arXiv:hep-ph/0006231]; T.~Fujita and K.~Ohashi,
  ``Superconformal tensor calculus in five dimensions,''
  Prog.\ Theor.\ Phys.\  {\bf 106}, 221 (2001)
  [arXiv:hep-th/0104130].
}

\lref\BergshoeffHC{
  E.~Bergshoeff, S.~Cucu, M.~Derix, T.~de Wit, R.~Halbersma and A.~Van Proeyen,
  ``Weyl multiplets of N = 2 conformal supergravity in five dimensions,''
  JHEP {\bf 0106}, 051 (2001)
  [arXiv:hep-th/0104113].
  ``N = 2 supergravity in five dimensions revisited,''
  Class.\ Quant.\ Grav.\  {\bf 21}, 3015 (2004)
  [Class.\ Quant.\ Grav.\  {\bf 23}, 7149 (2006)]
  [arXiv:hep-th/0403045].
}

\lref\KrausNB{
  P.~Kraus and F.~Larsen,
  ``Partition functions and elliptic genera from supergravity,''
  JHEP {\bf 0701}, 002 (2007)
  [arXiv:hep-th/0607138].
}


\Title{\vbox{\baselineskip12pt
}} {\vbox{\centerline {5D Attractors with Higher Derivatives}}}
\centerline{Alejandra Castro$^\dagger$\foot{aycastro@umich.edu},
Joshua L. Davis$^{\spadesuit}$\foot{davis@physics.ucla.edu}, Per
Kraus$^{\spadesuit}$\foot{pkraus@ucla.edu} and Finn
Larsen$^\dagger$\foot{larsenf@umich.edu}}
\bigskip
\centerline{${}^\dagger$\it{Michigan Center for Theoretical Physics,
Department of Physics}} \centerline{\it{University of Michigan, Ann
Arbor, MI 48109-1120, USA.}}\vskip.2cm
\centerline{${}^{\spadesuit}$\it{Department of Physics and
Astronomy, UCLA,}}\centerline{\it{ Los Angeles, CA 90095-1547,
USA.}}

\baselineskip15pt

\vskip .3in

\centerline{\bf Abstract}

We analyze higher derivative corrections to attractor geometries in
five dimensions.  We find corrected AdS$_3 \times S^2$ geometries by
solving the equations of motion coming from a recently constructed
four-derivative supergravity action in five dimensions.  The result
allows us to explicitly verify a previous anomaly based derivation
of the AdS$_3$ central charges of this theory.  Also, by dimensional
reduction we compare our results with those of the 4D higher
derivative attractor, and find complete agreement.

\Date{February, 2007}
\baselineskip14pt

\newsec{Introduction}

The last few years have seen progress in our understanding of
corrections to the entropy of black holes in string theory, both at
the microscopic and macroscopic levels.   On the supergravity side,
this has meant studying the effect of higher derivative terms in the
action \refs{\CardosoFP,\senrescaled,\us,\SenWA}. 4D extremal black
holes have a near horizon AdS$_2\times S^2$ geometry, with moduli
fixed by the attractor mechanism \attract.  By using the corrected
attractor solution and the general Wald entropy formula \wald, it is
possible to successfully match an infinite series of corrections to
the Bekenstein-Hawking area law with the corresponding microscopic
degeneracy of states
\refs{\CardosoFP,\senrescaled,\us,\OSV,\DabholkarYR,\DDMP,\KrausNB,\moreOSV,
\GaiottoNS}.  For reviews see \refs{\MohauptMJ,\PiolineNI,\KrausWN}.

However, on closer inspection this success actually seems quite
mysterious, since only a selected subset of terms in the
supergravity action are being used.  Namely, one incorporates the
supersymmetric completion of certain $R^2$ terms (as can be captured
by corrections to the generalized prepotential), but neglects
various $R^4$ and higher order terms, even though these {\it a
priori} contribute at the order one is working.   There is at
present no 4D understanding of why these terms can be neglected.

Greater control is achieved by realizing that these black holes
admit near horizon AdS$_3 \times S^2$ geometries \us. To relate an
AdS$_2 \times S^2$ geometry to AdS$_3 \times S^2$, one interprets
one of the 4D gauge fields as coming from a Kaluza-Klein circle
\refs{\StromBTZ}. An AdS$_3 \times S^2$ region then appears provided
that there is vanishing Kaluza-Klein monopole charge ($p^0=0$).
Alternatively, one can study these black holes in the context of 5D
supergravity.

By using the extra symmetries inherent in the 5D near horizon
description, one finds that the corrected entropy formula is
governed by the coefficients of the Chern-Simons terms in the
supergravity action.  In this way it is possible to bypass the need
to find the full set of higher derivative terms, or to find the
explicit values of the near horizon moduli.  The key observation is
that the entropy formula is controlled by the values of the left and
right moving central charges of the associated $1+1$ dimensional
CFT, and due to supersymmetry these are completely determined by
gauge and gravitational anomalies.

To verify this picture explicitly, and to find the  corrected black
hole geometry, one needs to work with the full 5D susy invariant
four derivative action, corresponding to the supersymmetric
completion of the four derivative Chern-Simons terms.   This action
appeared recently in \HanakiPJ.   In this paper we find the near
horizon AdS$_3 \times S^2$ geometry by analyzing the BPS conditions
and the equations of motion coming from this action. To do this, it
is most efficient to employ the ``$c$-extremization" procedure
developed in \us\ (or the closely related ``entropy-function"
developed in \SenWA). The strategy is to write down a c-function
whose critical points correspond to the solutions of the equations
of motion. Furthermore, the value of the c-function at a critical
point is equal to the average of the left and right moving central
charges of the associated CFT. We will show that the result is in
precise agreement with the values inferred from the
supersymmetry/anomaly based argument, and thereby verify that the
entropy is indeed controlled by the Chern-Simons terms. More
generally, this same logic leads to the conclusion that there are no
further corrections to the central charges from additional higher
derivative terms (i.e. more than four derivatives), since we have
already taken into account the full set of terms related by
supersymmetry to the Chern-Simons terms.

Since this procedure also yields the values of the fixed 5D moduli,
we can compare with the known results for the 4D moduli. Writing out
the details of the reduction from 5D to 4D we find full agreement.

This article is organized as follows. In section 2 we review
c-extremization. In section 3 we illustrate the procedure for the
leading order action. In section 4 we analyze the supersymmetry
conditions in an off-shell form that applies also when higher
derivatives are taken into account. In section 5 we carry out the
c-extremization procedure on the full action including all terms
that are related to the Chern-Simons term by supersymmetry. We find
results that are consistent with the supersymmetry conditions from
section 4 and moreover find a central charge that agrees with the
one previously found using supersymmetry and anomalies. In section 6
we compare our results with those found for black holes in four
dimensions and find complete agreement.

\newsec{Review of c-extremization}

The problem of finding an AdS$_3\times S^2$ solution\foot{Or more
generally an AdS$_p\times S^q$ solution} to a general higher
derivative action can be reduced to the problem of extremizing a
single function of the scale sizes and moduli \us.  Furthermore, the
value of this function at its critical point is (after suitable
normalization) equal to the average of the left and right moving
central charges of the asymptotic conformal symmetry group of the
theory.  In this section we review how this works.

We look for a solution respecting all AdS$_3 \times S^2$ isometries.
Besides constant scalar fields, we can also have two-form fields
proportional to the volume form on $S^2$.  More generally, we could
also have three-form fields proportional to the AdS$_3$ volume form;
these will not make any appearance in this paper, but we note that
they would necessitate a modification of some of the following
formulas.

Let the action for the theory be of the form
\eqn\aa{S= {1 \over 4\pi^2} \int\! d^5 x \sqrt{g}{\cal L} +S_{CS}+
S_{\rm bndy} ~,}
with $G_5={\pi\over4}$. Our trial solution takes the form
\eqn\ab{\eqalign{ds^2& = \ell_A^2  ds^2_{AdS} + \ell_S^2 d\Omega_2^2
\cr F^I&= {p^I \over 2}\epsilon_2 \cr v&= V \epsilon_2\cr    \phi^a
& = {\rm constant}~. }}
Here $F^I$ denote two-form fields strengths with magnetic charges
$p^I$; $v$ denotes additional two-form field(s); and $\phi^a$ denote
physical and auxiliary scalar fields. $\epsilon_2$ is the volume
form on the unit $S^2$. We normalize the $F^I$ such that the charges
$p^I$ are integer quantized.

 Since all covariant derivatives are assumed to vanish, the
equations of motion following from \aa, evaluated on the trial
solution \ab, reduce to extremizing the function $\sqrt{g} {\cal L}$
as a function of $\ell_A, \ell_S, V$ and $\phi^a$. Equivalently, we
can extremise the c-function, defined as
\eqn\ac{c= -6\ell_A^3\ell_S^2{\cal L}~. }
When extremizing, we hold fixed the quantized charges $p^I$, so that
all free parameters are determined in terms of the $p^I$ (or else
are undetermined).   This is the attractor mechanism, fixing the
geometry and  moduli in terms of the charges.

The  choice of normalization in \ac\ is motivated as follows. The
theory on AdS$_3$ has a boundary stress tensor \balkraus\ whose
trace anomaly  is \hensken\
\eqn\ad{ T^i_i = -{c \over 12}R^{(2)}~,}
where $R^{(2)}$ is the scalar curvature of the conformal boundary
metric.   The prefactor in \ac\ was chosen such that the c-function
evaluated at its critical point is equal to the $c$ appearing in
\ad.  In a theory with equal left and right moving central charges,
$c$ is the central charge.  More generally, the trace anomaly is
related to the average:
\eqn\ae{ c = \half(c_L + c_R)~. }

Given the central charges, evaluation of the Euclidean black hole
action leads to the general formula for the  black hole entropy $s$:
\eqn\af{s = 2\pi \sqrt{ {c_L \over 6}(L_0 - {c_L \over 24})} + 2\pi
\sqrt{ {c_R \over 6}(\tilde{L}_0 - {c_R \over 24})}~~.}
Since this expression takes the same form as Cardy's formula it is
convenient for comparison with microscopic results. Here, however,
it is just a statement about the on-shell supergravity action. In
fact, for $c_L= c_R$ \af\ agrees with Wald's entropy formula
independently of the microscopic theory
\refs{\saidasoda,\us}\foot{ Theories with $c_L \neq c_R$ have
gravitational Chern-Simons terms that violate diffeomorphism
invariance so Wald's formula does not apply. \TachikawaSZ\
generalizes Wald's formula to this case and shows that agreement
with \af\ is maintained. }.

We will be working with five dimensional supergravity coupled to
vector multiplets, which can be thought of as arising from
M-theory compactified on $CY_3$.  The charges $p^I$ then
correspond to M5-branes wrapping 4-cycles in $CY_3$.  The central
charges are known to be \refs{\MSW,\HMM}
\eqn\ag{ c_L= 6p^3 + \half c_2 \cdot p~,\quad c_R= 6p^3 +  c_2 \cdot
p~,}with \eqn\aga{p^3 = {1\over 6}c_{IJK}p^I p^J p^K~,}
where $c_{IJK}$ are the triple intersection numbers of the $CY_3$,
and $c_{2I}$ are the expansion coefficients of the second Chern
class.     This then gives the following prediction for the extremal
value of the c-function
\eqn\ah{ c= 6p^3 + {3 \over 4} c_2 \cdot p~.}
This is the result we wish to verify from the explicit higher
derivative action.

\newsec{ Two derivative analysis}

\subsec{Five dimensional off-shell supergravity.}

Following \HanakiPJ\ (see \refs{\KugoHN,\BergshoeffHC} for earlier
work) we consider superconformal gravity in five dimensions. The
local form of the theory is off-shell, meaning that the auxiliary
fields in the multiplets are not integrated out. At two-derivative
order the bosonic terms in the Lagrangian are\foot{We have omitted
fields in the multiplets associated with gauged supergravity or that
can be turned off by gauge fixing conformal symmetries. With respect
to \HanakiPJ\ we have switched the sign in the kinetic term for the
scalars $M^I$ and the sign of the Ricci scalar
($R_{here}=-R_{there}$).} \eqn\ba{\eqalign{{1\over2}{\cal L}_0=&
\partial^a{\cal A}_i^{\alpha}\partial_a{\cal
A}^i_{\alpha}+{\cal
A}^2\left({1\over8}D-{3\over16}R-{1\over4}v^2\right)\cr &+{\cal
N}\left({1\over4}D+{1\over8}R+{3\over2}v^2\right)+{\cal
N}_Iv^{ab}F^I_{ab}\cr&+{\cal
N}_{IJ}\left({1\over8}F^I_{ab}F^{Jab}+{1\over4}\partial_aM^I\partial^aM^J\right)+{1\over48}e^{-1}
c_{IJK}A^I_{a}F^{J}_{bc}F^{K}_{de}\epsilon^{abcde}~.}}

We are taking into account the bosonic fields of two distinct super
multiplets: the Weyl multiplet, contains the vielbein
$e_\mu^{\phantom{\mu}a}$, the two-form auxiliary field $v_{ab}$, and
a scalar auxiliary field $D$; the vector multiplets enumerated by
index $I=1\ldots n_V$, each containing a one-form gauge field $A^I$
and scalar $M^I$, with $F^I=dA^I$. Although we will not discuss
gauge fixing in detail, it is useful to include a term for the hyper
multiplet which contains the Weyl scalar ${\cal A}^\alpha_i$. The
index $i=1,2$ refers to $SU(2)$ doublets and $\alpha=1,\ldots 2r$
refers to the $USp(2r)$. The hyper is used to gauge fix the
dilatational symmetry and we choose a gauge that satisfies
\eqn\bca{{\cal A}^2=-2~, \quad
\partial_a{\cal A}^i_\alpha=0~.}

The functions on the scalar manifold  are defined by\eqn\bc{{\cal
N}={1\over6}c_{IJK}M^IM^JM^K~,\quad {\cal N}_I=\partial_I{\cal
N}={1\over2}c_{IJK}M^JM^K~,\quad {\cal N}_{IJ}=c_{IJK}M^K~.}
The auxiliary field $D$ appears linearly in \ba, which means that
it acts as a Lagrange multiplier. The resulting constraint
determines ${\cal N}$, which can be thought of as the volume of
the compactification manifold. Given that we chose ${\cal
A}^2=-2$, solving the equation of motion for $D$ implies ${\cal
N}=1$. So, at the level of the two derivative action the scalars
are described using real special geometry. For a pedagogical
introduction see \LarsenXM.

We can eliminate the auxiliary fields $v_{ab}$ and $D$ by solving
their equations of motion. This gives

\eqn\bb{{\cal L}_0=-{\cal N}
\bigg[-R+G_{IJ}\partial_aM^I\partial^aM^J+{1\over2}G_{IJ}F^I_{ab}F^{Jab}-{e^{-1}\over24{\cal
N}}c_{IJK} A^I_{a}F^{J}_{bc}F^{K}_{de}\epsilon^{abcde}\bigg]~,} with

\eqn\bd{G_{IJ}=-{1\over2}\partial_I\partial_J(\ln{\cal
N})={1\over2}\left({ {\cal N}_I{\cal N}_J\over{\cal N}^2}-{{\cal
N}_{IJ}\over{\cal N}}\right)~.}
This is the familiar two derivative Lagrangian in five dimensional
supergravity. For our purposes, we will not use \bb\ and instead
work with \ba.

\subsec{c-extremization}

We now determine, at the two-derivative order, the near horizon
AdS$_3 \times S^2$ geometry corresponding to a black string in five
dimensions, which we will refer to as the ``black string attractor".
The near horizon configuration is given by \ab\ and the central
charge as defined in \ac\ is

\eqn\dc{c=-6\ell_A^3\ell_S^2{\cal L}_0~,} with ${\cal L}_0$ given by
\ba\ evaluated on the trial solution \ab. For this configuration, we
will have some simplifications. The Chern-Simons term in \ba\
vanishes and derivatives of the scalars $M^I$ are zero. The Ricci
scalar is

\eqn\dca{R=-{6\over\ell_A^2}+{2\over\ell_S^2}~.}

By symmetry, we know that  the scalars $M^I$ are proportional to
the charges $p^I$, so we write $M^I=mp^I$, with $m$ a constant to
be determined. The $c$-function then becomes,

\eqn\dd{\eqalign{c=-12\ell_A^3\ell_S^2\bigg(&{1\over4}(p^3m^3-1)D-{1\over4}(p^3m^3+3)\left({3\over
\ell_A^2}-{1\over \ell_S^2}\right)\cr&+{1\over
\ell_S^4}\left((3p^3m^3+1)V^2+3p^3m^2V\right)+{3p^3\over
\ell_S^4}{m\over8}\bigg)~,}} with $p^3$ given by equation \aga.
Extremizing \dd\ with respect to $D$ imposes $m^3=p^{-3}$ and the
equation for $V$ gives

\eqn\dda{V=-{3\over 8}p~.}
The extremization of \dd\ with respect to the radii $\ell_A$ and
$\ell_S$ results in

\eqn\ddb{\ell_A=2\ell_S~,\quad \ell_A=p~.}

Finally, by extremizing the $c$-function with respect to $m$ we find
$D=12p^{-2}$. Summarizing, our result for the parameters of the
solution is

\eqn\de{M^I={p^I\over p}~, \quad \ell_A=2\ell_S=p~, \quad D={12\over
p^2}~,\quad V=-{3p\over8}~.} Inserting \de\ in \dd, the central
charge for the black string in the two derivative theory is

\eqn\df{c=6p^3=c_{IJK}p^Ip^Jp^K~.}

The value of $c$  agrees with the expectation \ah\ to the leading
order in charges. The new feature is verifying that
$c$-extremization off-shell (i.e. keeping auxiliary fields) is
consistent.

\newsec{Susy variations}

We would like to determine corrections to the attractor solution
from the higher derivative terms in the action.  A strong
constraint comes from the fact that the attractor solution
exhibits maximal supersymmetry.  Furthermore, in the off-shell
formulation the supersymmetry transformations are independent of
the detailed form of the action (i.e. they are the same for the
two and four derivative actions).   With this in mind, we now
analyze the constraints from supersymmetry.

The supersymmetry variations are
 \eqn\aaa{\eqalign{\delta\psi^i_\mu&={\cal
D}_\mu\varepsilon^i+{1\over2}v^{ab}\gamma_{\mu
ab}\varepsilon^i-\gamma_\mu\eta^i~, \cr \delta
\chi^i&=D\varepsilon^i-2\gamma^c\gamma^{ab}\varepsilon^i{\cal
D}_av_{bc}-2\gamma^a\varepsilon^i\epsilon_{abcde}v^{bc}v^{de}+
4\gamma\cdot v\eta^i~, \cr
\delta\Omega^{Ii}&=-{1\over4}\gamma\cdot
F^I\varepsilon^i-{1\over2}\gamma^a\partial_aM^I\varepsilon^i-M^I\eta^i~,
\cr\delta\zeta^\alpha&=\gamma^a\partial_a{\cal
A}^\alpha_j\varepsilon^j-\gamma\cdot v\varepsilon^j{\cal
A}^\alpha_j+3{\cal A}^\alpha_j\eta^j~.}}

The first two transformations come from the fermions in the Weyl
multiplet, the gravitino $\psi^i_\mu$ and an auxiliary Majorana
spinor $\chi^i$. From the vector multiplets we have the gaugino
$\Omega^I_i$ and the hyper multiplet contributes with
$\zeta^\alpha$. We are using the notation $\gamma\cdot
v=\gamma_{ab}v^{ab}$.

\subsec{Supersymmetry constraints for the black string attractor.}

The supersymmetry transformations \aaa\ simplify dramatically when
evaluated on our trial background \ab. The attractor has maximal
supersymmetry, meaning all variations must vanish. In our background
this reduces to solving

\eqn\ea{\eqalign{\delta\psi^i_\mu&={\cal
D}_\mu\varepsilon^i+{1\over2}v^{ab}\gamma_{\mu
ab}\varepsilon^i-\gamma_\mu\eta^i=0~, \cr \delta
\chi&=D\varepsilon^i+ 4\gamma\cdot v\eta^i=0~, \cr
\delta\Omega^{Ii}&=-{1\over4}\gamma\cdot
F^I\varepsilon^i-M^I\eta^i=0~,\cr \delta\zeta^\alpha&=(-\gamma\cdot
v\varepsilon^j+3\eta^j){\cal A}^\alpha_j=0~.}}
From the gaugino variation it is clear that the scalars $M^I$ are
proportional to the charges $p^I$, so we can write $M^I=mp^I$, where
the constant of proportionality $m$ will be determined by the
remaining equations. The last equation in \ea\ gives

\eqn\eab{\eta^i={1\over3}\gamma\cdot v\varepsilon^i~.}
Inserting \eab\ in \ea\ we get for the gravitino variation

\eqn\eac{\left({\cal D}_\mu+{1\over2}v^{ab}\gamma_{\mu
ab}-{1\over3}v^{ab}\gamma_\mu\gamma_{ab}\right)\varepsilon^i=0~,}
and for the auxiliary field and gaugino  \eqn\eae{\eqalign{\left(D+
{4\over 3}(\gamma\cdot v)^2\right)\varepsilon^i&=0~,\cr
\left(-{1\over4}\gamma\cdot F^I-{m\over3}p^I\gamma\cdot
v\right)\varepsilon^i&=0~.}}
Solving \eae\ on the ansatz \ab, we find

\eqn\eb{mV=-{3\over8}~,\quad D={16\over3}{V^2\over \ell_S^4}~.}

Finally, from the gravitino variation \eac\foot{See \ChamseddinePI,
\GutowskiYV\ and references therein for details on manipulation of
the gravitino variation.} we get a relation between the radii,
$\ell_A$ and $\ell_S$, and the auxiliary field $V$

\eqn\ec{\ell_A=2\ell_S~, \quad V=-{3\over8}\ell_A~.}

The relations between the moduli shown in equations \eb\ and \ec\
hold independently of the action. Since the supersymmetry variations
are exact off-shell, these results will not change for higher
derivatives theories.

It is important to note that supersymmetry does not fully determine
the values of the moduli and as presented here, one of the fields is
unconstrained. This should be expected, since there are gauge
symmetries unrelated to supersymmetry transformations that we have
not imposed . For example, in the leading order theory described by
\ba, the scalars are described using real special geometry, where
the volume ${\cal N}$ is fixed. This comes about from fixing the
superconformal theory to Poincare supergravity using the equation of
motion for $D$, and is not related to the fact that the theory is
supersymmetric. When higher derivatives terms are included in the
theory, one should similarly expect to use at least one equation of
motion from the off-shell theory to specify the solution completely.

Without loss of generality, we will take the $AdS_3$ radius as the
undetermined modulus, and so we can summarize our results as

\eqn\ed{\ell_S={1\over2}\ell_A~,\quad m={1\over \ell_A}~,\quad
V=-{3\over8}\ell_A~,\quad D={12\over \ell_A^2}~.}

Comparing with the two derivative $c$-extremization, we can see that
\ed\ agrees with \de. The piece of information that is missing from
the supersymmetry constraints is the relation between $\ell_A$ and
the charges $p^I$. At the level of the two derivative theory, this
is simply $\ell_A=p$. As we will show in the next section, when
higher derivatives are taken into account, the $AdS_3$ radius will
be modified and the value of the corrected moduli will be determined
by our procedure.

\newsec{$c$-extremization including higher derivatives}

We are now ready to discuss higher derivative corrections to the
central charge. As mentioned in the introduction, we want to verify
that the Chern-Simons term controls the corrections to the central
charge. From anomaly arguments, this term is given by

\eqn\cb{\sqrt{g}{\cal L}_{CS}= -{c_{2I}\over 48\cdot
2}A^I\wedge\hbox{Tr}(R\wedge R)={c_{2I}\over 24\cdot
16}\epsilon_{abcde}A^{Ia}R^{bcfg}R^{de}_{\phantom{de}fg}~.}
The Chern-Simons term by itself is not supersymmetric and therefore
extra terms should be included. The four derivative supersymmetric
completion of \cb\ was computed in \HanakiPJ, and the relevant terms
for our discussion are

\eqn\ca{\eqalign{{\cal L}_1={c_{2I}\over
24}\bigg(&{1\over16}e^{-1}\epsilon_{abcde}A^{Ia}C^{bcfg}C^{de}_{\phantom{de}fg}+{1\over8}M^IC^{abcd}C_{abcd}+{1\over12}M^ID^2
+{1\over6}F^{Iab}v_{ab}D\cr &-{1\over3}M^IC_{abcd}v^{ab}v^{cd}
-{1\over2}F^{Iab}C_{abcd}v^{cd}+{8\over3}M^{I}v_{ab}\hat{{\cal
D}}^b\hat{{\cal D}}_cv^{ac}\cr& +{4\over3}M^I\hat{{\cal
D}}^av^{bc}\hat{{\cal D}}_av_{bc}+{4\over3}M^I\hat{{\cal
D}}^av^{bc}\hat{{\cal
D}}_bv_{ca}-{2\over3}e^{-1}M^I\epsilon_{abcde}v^{ab}v^{cd}\hat{{\cal
D}}_fv^{ef}\cr &
+{2\over3}e^{-1}F^{Iab}\epsilon_{abcde}v^{cd}\hat{{\cal
D}}_fv^{ef}+e^{-1}F^{Iab}\epsilon_{abcde}v^c_{\phantom{c}f}\hat{{\cal
D}}^dv^{ef} \cr & -{4\over3}F^{Iab}v_{ac}v^{cd}v_{db}
-{1\over3}F^{Iab}v_{ab}v^2+4M^Iv_{ab}v^{bc}v_{cd}v^{da}-
M^I(v_{ab}v^{ab})^2\bigg)~,}} with $C_{abcd}$ the Weyl tensor
defined as

\eqn\caa{C^{ab}_{\phantom{ab}cd}=R^{ab}_{\phantom{ab}cd}+{1\over6}R\delta^{[a}_{\phantom{a}[c}\delta^{b]}_{\phantom{b}d]}
-{4\over3}\delta^{[a}_{\phantom{a}[c}R^{b]}_{\phantom{b}d]}~.}
The double covariant derivative of $v_{ab}$ has curvature
contributions\foot{The sign difference with respect to \HanakiPJ\
is coming from the difference in curvature convention.} given by
\eqn\cab{v_{ab}\hat{{\cal D}}^b\hat{{\cal D}}_cv^{ac}=v_{ab}{\cal
D}^b{\cal
D}_cv^{ac}+{2\over3}v^{ac}v_{cb}R_{a}^{\phantom{a}b}+{1\over12}v_{ab}v^{ab}R~.}

\subsec{Central Charge and Moduli Corrections}

Using the $c$-extremization procedure explained in section 2, we
will find the corrected central charge and moduli in the higher
derivative theory for the $5D$ black string. Including the four
derivative Lagrangian, the central charge becomes
\eqn\fb{c=-6\ell_A^3\ell_S^2({\cal L}_0+{\cal L}_1)~,} where ${\cal
L}_0$ and ${\cal L}_1$ are given by \ba\ and \ca\ evaluated in the
$AdS_3\times S^2$ background. Using \ab\ and the attractor value for
the moduli $M^I=mp^I$, the supersymmetric four derivative
contributions to the central charge are \eqn\fd{\eqalign{{\cal
L}_1={c_2\cdot p\over 24}\bigg[&{m\over4}\left({1\over
\ell_A^2}-{1\over \ell_S^2}\right)^2+{2\over3}{V^3\over
\ell_S^8}+4m{V^4\over \ell_S^8}+m{D^2\over12}+{D\over6}{V\over
\ell_S^4}\cr& -{2\over3}m{V^2\over \ell_S^4}\left({3\over
\ell_A^2}+{5\over \ell_S^2}\right)+{1\over2}{V\over
\ell_S^4}\left({1\over \ell_A^2}-{1\over \ell_S^2}\right)\bigg]~.}}
On our trial background $\hat{\cal D}_av_{bc}=0$. The two derivative
contribution to \fb\ is still given by \dd.

According to $c$-extremization we should now extremize with respect
to all parameters. It would be extremely difficult to do this were
it not for the guidance provided by the supersymmetry analysis in
the previous section. For example, the variation of \fb\ with
respect to $m$ gives
\eqn\fe{\eqalign{&{3p^3m^2\over4}\left(D-{3\over \ell_A^2}+{1\over
\ell_S^2}\right)+{3p^3\over \ell_S^4}\left(3m^2V^2+2mV+{1\over
8}\right)+\cr &+{c_2\cdot p\over 48}\left[{1\over4}\left({1\over
\ell_A^2}-{1\over \ell_S^2}\right)^2+4{V^4\over
\ell_S^8}+{D^2\over12}-{2\over3}{V^2\over \ell_S^4}\left({3\over
\ell_A^2}+{5\over \ell_S^2}\right)\right]=0~.}}
It is easy to verify that the moduli given in \ed\ do indeed satisfy
this equation.

Before proceeding with the remaining extremization conditions,
recall that the BPS conditions \ea\  leave one modulus undetermined.
The missing information about the attractor can be simply obtained
from the $c$-extremization procedure. The simplest is to consider
the equation for the auxiliary field $D$
\eqn\ff{{\partial c\over \partial D}=0\quad \Rightarrow \quad
p^3m^3=1-{1\over72}c_2\cdot p\left(mD+{V\over \ell_S^4}\right)~.}
Using \ed\ and solving for the $AdS_3$ radius ($\ell_A$) we find
\eqn\fg{\ell_A^3=p^3+{1\over12}c_2\cdot p~.}
The solution is now fully specified. At this point it is
straightforward to vary the c-function \fb\ also with respect to
$V$, $\ell_A$, $\ell_S$ and show that the resulting equations are
satisfied by \ed\ and \fg. All in all we have found a solution
that extremizes the c-function and shown that this solution is
supersymmetric. Since we proceeded somewhat indirectly, we have
not excluded the existence of other solutions with the same charge
configuration but no supersymmetry.

Evaluating the c-function for our solution we find the corrected
central charge
\eqn\fh{c=6p^3+{3\over4}c_2\cdot p~.}
This is precisely the result \ah\ that was predicted from
supersymmetry and anomalies.

It is worth noting that the simple form of this result comes about
in a rather nontrivial way in the present procedure. The radius of
curvature $\ell_A$ \fg\ is a nontrivial function of the charges and
it enters in the denominator of the Lagrangian \fd. It is only due
to intricate cancellations that the final result \fh\ becomes a
linear function of the charges $p^I$. That this works out correctly
provides a rather stringent consistency check on the entire
framework.

\subsec{Small black holes}

One of the benefits of considering higher derivative corrections is
that we can find smooth solutions in cases where the two-derivative
action would yield a naked singularity. These are so-called ``small
black holes" \refs{\DabholkarYR,\DDMP}. In particular, if we choose
charges $p^I$ such that $p^3=0$ but $c_2 \cdot p \neq 0$, then the
two-derivatives formulas \de\ become singular, while the four
derivative formulas given in this section are well behaved. We can
further check that the full action expanded around the small black
hole solution exhibits no obvious pathologies.   On the other hand,
since some of the moduli $M^I$ now vanish, some of the internal
compactification cycles are becoming small, and so one should be
alert to potentially large corrections from non-perturbative effects
not included here. See \DDMP\ for more discussion.

\newsec{Comparison between 5D and 4D attractor formulas}

So far we have focussed on the $AdS_3\times S^2$ attractor geometry
near an effective string in five dimensions. Most recent works on
higher curvature corrections are in the context of black holes in
four dimensions and their $AdS_2\times S^2$ near horizon attractor
geometry. It is instructive to work out the detailed comparison
between the four and five dimensional perspective in view of the
higher derivative corrections.  This is achieved by wrapping the
string on a circle and dimensionally reducing. In terms of black
hole entropy counting, a recent discussion of the relation between
the AdS$_3$ and AdS$_2$ viewpoints can be found in \DabholkarTB.

A good starting point is the $AdS_3$ geometry written in Poincare
coordinates as
\eqn\ga{ ds^2_3 = {\ell^2_A\over y^2}(dw^+dw^- + dy^2)~. }
Introducing the coordinates \MaldacenaBW
\eqn\gb{\eqalign{ w^+ & = {1\over 2\pi T_L} e^{2\pi T_L(x^5 + t)~,}
\cr w^- & = x^5 -t - {\ell^4_A \pi T_L\over U^2}~,\cr y &=
{\ell^2_A\over U} e^{\pi T_L (x^5 + t)}~, }}
 the line element becomes
\eqn\gc{ ds^2_3 = {U^2\over\ell^2_A} (dx^2_5 - dt^2) + \ell^2_A
{dU^2\over U^2} + \pi^2 \ell^2_A T^2_L (dx_5 + dt)^2~. }
Wrapping the string on a circle corresponds to imposing the
periodicity condition $x^5 \equiv x^5 + 2\pi R$, which amounts to
identifications on the $AdS_3$ space that change the causal
structure to that of a black hole \BanadosGQ. This is clearest if we
introduce the Schwarzschild-type coordinates
\eqn\gd{\eqalign{ \rho^2 & = (\pi^2 \ell^2_A T^2_L +
{U^2\over\ell^2_A})R^2~, \cr x_5 &= R\phi~, \cr \tau &= {\ell_A\over
R}t~. }}
Then the line element becomes
\eqn\gea{ ds^2_{\rm BTZ} = - N^2 d\tau^2 + N^{-2}d\rho^2 +
\rho^2(d\phi + N_\phi d\tau)^2~, } where \eqn\gf{\eqalign{ N&  =
{\rho\over\ell_A} - {\pi^2  \ell_A T^2_L R^2\over\rho}~, \cr N_\phi
& = {\pi^2 \ell_A T^2_L R^2 \over\rho^2}~. }}
%

We are interested in the direct product of the three dimensional
geometry just introduced and an additional $S^2$. Kaluza-Klein
reduction on $x^5$ takes us from 5D to 4D. The compactification
yields a 4D dilaton with  near horizon value
\eqn\gh{ e^{-2\Phi} = \rho_{\rm hor} = \pi \ell_A T_L R~. }

So far we have just reviewed a standard construction. The issue we
wish to emphasize is that these considerations are purely geometric
and thus hold regardless of the details of the action. The
corrections due to higher curvature terms enter only through the
relation between parameters in the geometry and the underlying
microscopic parameters.  For example, we found earlier that the
$AdS_3$ radius is
\eqn\gha{ \ell_A = p_R~, } where \eqn\gl{ p_R^3 = {1\over 6} \left(
c_{IJK} p^I p^J p^K + {1\over 2}c_{2}\cdot p\right)~. }
We also found the attractor values of the 5D scalars as
\eqn\gla{ X^I = {p^I\over p_R}~. }
Note that the corrections again appear through $p_R$.

We are also interested in the  thermodynamics for a black string
excited to level $|q_0|$. The corrected formula for the entropy is
\eqn\gm{ S = 2\pi \sqrt{c_L |q_0|\over 6} = 2\pi\sqrt{p_L^3 |q_0|}~,
} where the corrections enter through \eqn\glb{\eqalign{ p_L^3 =
{1\over 6} \left( c_{IJK} p^I p^J p^K + c_{2}\cdot p\right)~.\cr}}
The energy of the excitations is $E_L = |q_0|/R$ and so the first
law of thermodynamics gives the temperature \eqn\gn{ T_L= {1\over\pi
R}\sqrt{|q_0|\over p_L^3}~. } We see again that it is the
combination $p_L$ that appears in the thermodynamics.

We next consider the corrections of some less obvious quantities.
For example, the precise value for the 4D dilaton \gh\ is \eqn\go{
e^{-2\Phi} = p_R \sqrt{|q_0|\over p_L^3}~. } The string frame radius
of the very near horizon $AdS_2$ close to the 4D black hole is
inherited from the $AdS_3$ \StromingerYG\ and so its value is simply
$\ell_A$. The $AdS_2$ radius in 4D Einstein frame is therefore
\eqn\goa{ R_0^2 = e^{-2\Phi} \ell_A^2 = p_R^3 \sqrt{|q_0|\over
p_L^3}~. } This expression agrees with the result previously found
directly in four dimensions (for a good review see \MohauptMJ
\foot{The $AdS_2$ radius is the 4D central charge so $R^2_0=|Z|^2$.
Combining (6.16) and (6.17) in \MohauptMJ\ gives our \goa\ after
notation has been adapted.}).

We can also determine the 4D scalars. We are considering the simple
situation with $q_{I\neq 0}=0$ and $p^0=0$, where there are no
M2-branes wrapping the $2$-cycles of the CY, nor any magnetic charge
of the Kaluza-Klein gauge field from reduction along $x^5$. In this
case the 4D scalars other than the dilaton \go\ are purely
imaginary.\foot{M2-brane charges $q_I$ are easily incorporated, as
they correspond to Wilson lines for the gauge fields: $w^I \sim {1
\over R} A^I_5 \sim c^{IJ}q_J$, with $c^{IJ}$ being the inverse of
$c_{IJK}p^K$. } Combining the 5D scalars \gla\ with the dilaton we
find \BehrndtAC\  \eqn\gp{ z^I = ie^{-2\Phi} X^I =ip^I
\sqrt{|q_0|\over p_L^3}~. } This expression also agrees with results
previously found directly in four dimensions\foot{Adapting the
notation of (6.18) in \MohauptMJ\ gives our \gp}. Note that in 4D
the $p_L$ introduced in \glb\ controls both the scalars and the
thermodynamics.

From the 5D point of view the charge $q_0$ corresponds to AdS$_3$
angular momentum.    In a two-derivative theory the angular momentum
can be read off from the metric via
\eqn\gq{ j = {\rho_{\rm hor}^2\over 4G_3\ell_A}~. }  Applying this
formula  to the corrected metric yields (in our units $G_3 = {1\over
4\ell^2_A}$)
\eqn\gr{j= {p_R^3\over p_L^3} |q_0|~.}
The mismatch between $j$ and $q_0$ is due to the fact that the
expressions for conserved quantities  as surface integrals
themselves receive corrections from the higher derivative terms. It
would be instructive to derive these corrected expressions, as was
done in \KrausZM\ for the gravitational Chern-Simons term.

In this section we have focussed on extremal black holes with
$T_L\neq 0$ and $T_R=0$.   However, one of the nice features of the
AdS$_3$ framework is that our 5D results easily extend to the
non-extremal case $T_{L,R}\neq 0$.  The 5D attractor formulas are
unchanged, and the general entropy formula is given in \af.  So
higher derivative corrections to the entropy are under control even
for these non-BPS, non-extremal black holes \us.

Another feature of the 5D setup is that we can make contact with
higher derivative corrections to black rings \rings.  Black rings
have a near horizon AdS$_3 \times S^2$ region of the same type as
studied here.  Therefore our results for the corrected attractor
geometry will also apply to the near horizon region of black rings.
In particular, it should be possible to find explicit solutions for
``small black rings".

\bigskip
\noindent {\bf Acknowledgments:} \medskip \noindent We thank K.
Hanaki for discussions.   Work of PK and JD  supported in part by
NSF grant PHY-0456200. The work of FL and AC is supported  by DOE
under grant DE-FG02-95ER40899.

\listrefs
\end